\documentclass{PoS}

\title{Nuclear Parton Distributions and the Drell-Yan Reaction}

\ShortTitle{Nuclear PDFs}

\author{S.A. Kulagin$^a$ and R. Petti$^b$ \\
        \llap{$^a$}Institute for Nuclear Research of the Russian Academy of Sciences, Moscow 117312, Russia\\
        \llap{$^b$}Department of Physics and Astronomy, University of South Carolina, Columbia SC 29208, USA\\
        E-mail: \email{kulagin@ms2.inr.ac.ru}, \email{roberto.petti@cern.ch}
        }

\abstract{%
We discuss the nuclear parton distribution functions on the basis of
our recently developed semi-microscopic model, 
which takes into account a number of nuclear effects including
Fermi motion and nuclear binding, nuclear meson-exchange currents and off-shell corrections
to bound nucleon distributions as well as nuclear shadowing effect.
We also discuss application to the nuclear Drell-Yan process and
compare our predictions with data from the E772 and E866 experiments.
}

\FullConference{XXIII International Workshop on Deep-Inelastic Scattering\\
		27 April - May 1 2015\\
		Dallas, Texas}

\begin{document}

\section{Introduction}

The parton distributions (PDFs) are universal process-independent characteristics
of the target at high invariant momentum transfer $Q$, which determine the leading contributions 
to the cross sections of various hard processes involving leptons and hadrons.
The nuclear parton distributions (nPDFs) are the subject of significant nuclear effects
which span a wide region of Bjorken $x$ with a rate more than 
one order of magnitude larger than the ratio of the nuclear 
binding energy to the nucleon mass (for a review see \cite{Arneodo:1992wf,Norton:2003cb}).
These observations indicate that the nuclear environment plays an important role
even at energies and momenta much higher 
than those involved in typical nuclear ground state processes.

A few phenomenological approaches to nPDFs are available in literature
\cite{Eskola:2009uj,Hirai:2007sx,deFlorian:2011fp,Kovarik:2015cma}.
Although such studies are useful in constraining nuclear effects
for different partons, they provide little information about the underlying physics mechanisms 
responsible for the nuclear corrections.
In this contribution we follow a different approach and present the results of a study of nPDFs
using a semi-microscopic model of Ref.\cite{KP04}.
The model accounts for a number of nuclear corrections including the smearing with the
energy-momentum distribution of bound nucleons (Fermi motion and binding, or FMB),
the off-shell correction (OS) to bound nucleon structure functions, 
the contributions from meson exchange currents (MEC), and the effects of propagation of 
the hadronic component of the virtual intermediate boson in the nuclear environment.
The model quantitatively explains the observed $x$, $Q^2$ and $A$ dependencies of
the nuclear structure functions in the deep-inelastic scattering (DIS) 
for a wide range of nuclear targets from $^2$H to ${}^{207}$Pb~\cite{KP04,KP07,KP10}.
Here we discuss the region of high $Q$ and address the predictions of the model for nPDFs
and the nuclear Drell-Yan (DY) process
(see Ref.\cite{KP14} for more detail).

\section{Outline of a model}
\label{model}

We will use the notation $q_{a/T}(x,Q^2)$ for the distribution of quarks of the flavor
$a$ in a target $T$. The parton distributions in a nucleus
receive a number of contributions and can be written as \cite{KP04,KP14}
(for brevity, we suppress explicit dependencies on $x$ and $Q^2$): 
\begin{equation}
\label{npdf}
q_{a/A} = \left\langle q_{a/p}\right\rangle + \left\langle q_{a/n}\right\rangle 
          + \delta q_a^\mathrm{coh} + \delta q_a^\mathrm{MEC} .
\end{equation}
The first two terms on the right side are the contribution from the partons from
bound protons and neutrons averaged with the nuclear spectral function,
as discussed in detail in Ref.\cite{KP04,KP14}.
Note that the evaluation of these terms requires the proton and the neutron PDFs 
in off-mass shell region \cite{Kulagin:1994fz}.
The off-shell correction together with the nucleon momentum distribution (Fermi motion)
and the nuclear binding effect~\cite{FMB}
plays an important role in the valence quark region \cite{KP04,KP10}.

The $\delta q^\mathrm{coh}$ correction is relevant at low $x$ and
arises due to propagation effects of intermediate hadronic states
of a virtual boson in nuclear environment.
This term involves contributions from multiple scattering series
and typically lead to a negative correction (nuclear shadowing effect,
for a review see, e.g., Ref.\cite{Piller:1999wx}).

The last term in Eq.(\ref{npdf}) is a contribution from the pion (as well as the other meson)
degrees of freedom in nuclei.
This term is relevant for intermediate region of $x$ and results in a some enhancement
of the nuclear sea quark distribution. Also its contribution is important to balance the overall
nuclear light-cone momentum. 
More details on the treatment of each term in Eq.(\ref{npdf}) can be found
in Refs.\cite{KP04,KP14}.
Below we summarize our results on the nuclear valence and sea quark PDFs.

\section{Results}
\label{res}

\begin{figure}[htb]
\centering
\includegraphics[width=0.5\textwidth]{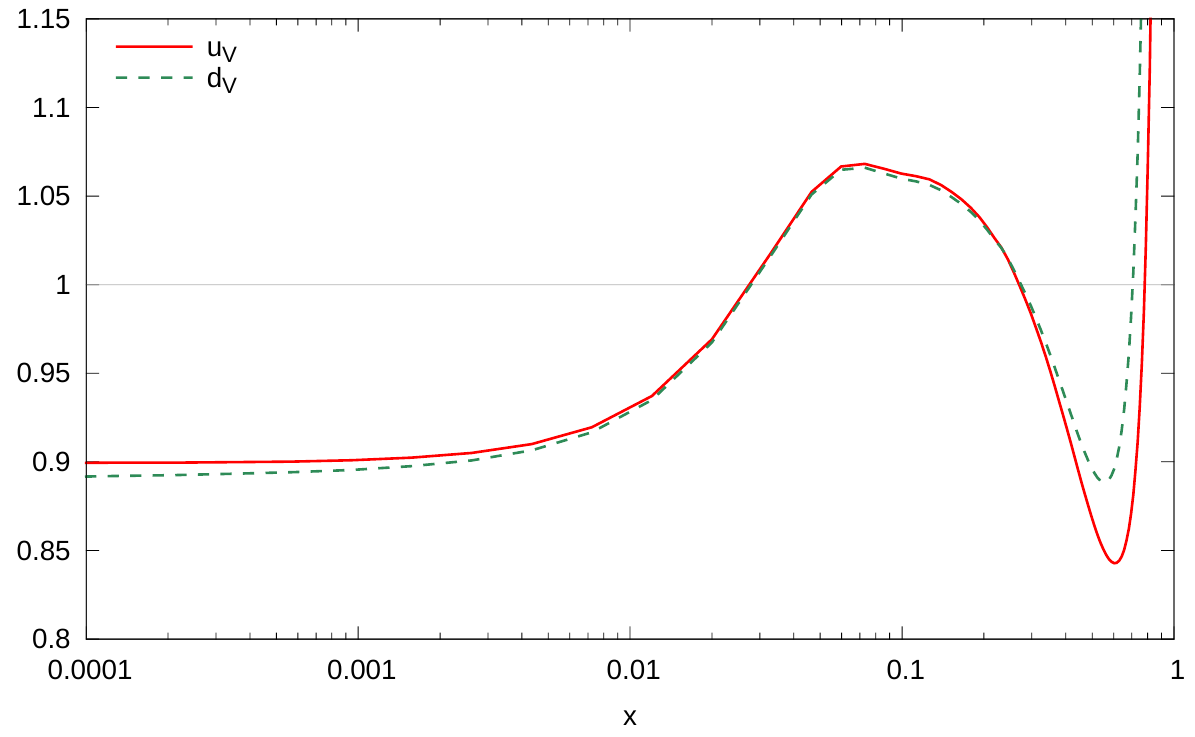}%
\includegraphics[width=0.5\textwidth]{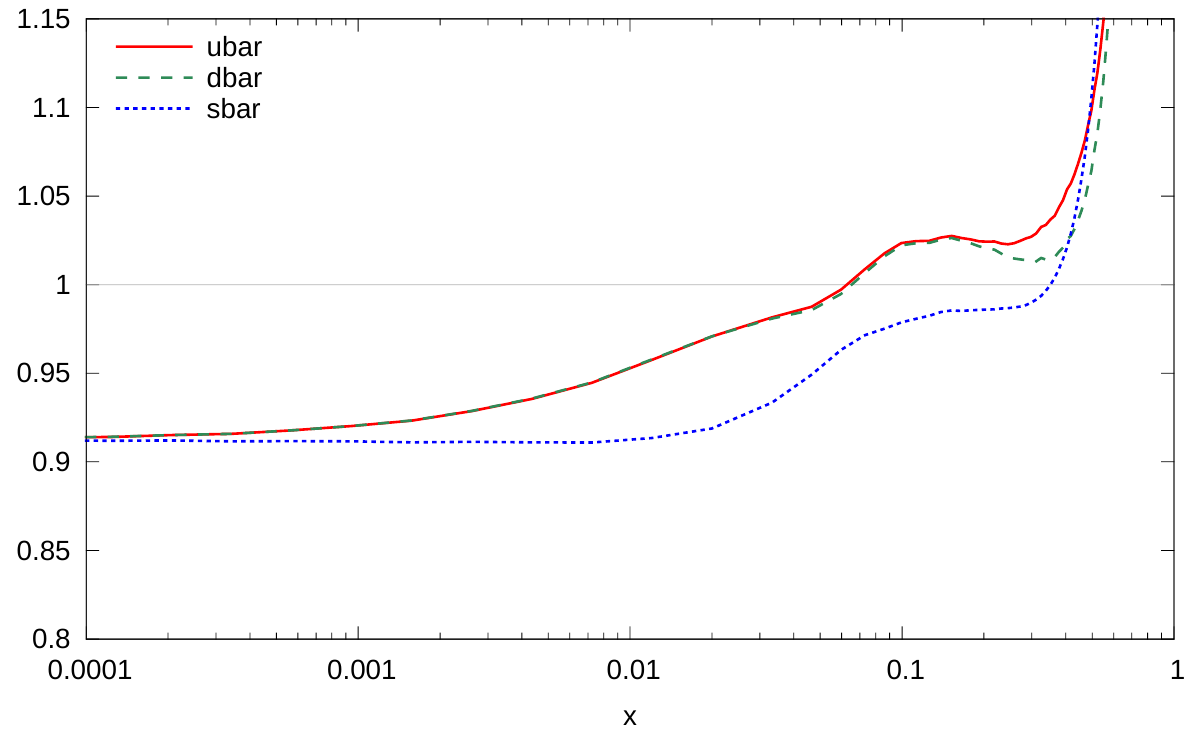}
\caption{%
(Color online)
The nuclear ratios for the valence quarks (left panel) and the $\bar u$, $\bar d$, and $\bar s$ antiquarks (right panel) in ${}^{208}$Pb calculated at $Q^2=25\, \mathrm{GeV}^2$
(see text for more detail).
\label{fig:npdfs}}
\end{figure}

Figure~\ref{fig:npdfs} displays
the ratios $R_a=q_{a/A}/(Z q_{a/p}+N q_{a/n})$
for the valence quarks (left panel) and the antiquarks (right panel) in the lead nucleus.
In the definition of this ratio, $Z$ and $N$ are the proton and the neutron number,
and $q_{a/p}$ and $q_{a/n}$ are the free proton and the neutron PDF, respectively.
As can be seen the resulting nuclear effect is different for the nuclear valence
and sea distribution and also depends on the PDF flavor.
We briefly comment on characteristic features of the nuclear ratios for different PDFs,
for a more detail discussion see Ref.\cite{KP14}.

In the region $x\ll 0.1$, all nPDFs are suppressed
due to the nuclear \emph{shadowing} effect (the $\delta q^{\rm coh}$ term).
Note that this effect is not universal and differs for
the valence and sea quark distribution \cite{KP04,KP14}.%
\footnote{%
In our analysis we choose the PDFs with definite $C$-parity
$q^\pm = q \pm \bar q$ as a basis.
The $C$-odd $xq^-$ and the $C$-even $xq^+$ distributions correspond to the structure function 
$xF_3$ and $F_2$, respectively.
For  light quarks we also consider the isoscalar $q_0=u+d$ and the isovector $q_1=u-d$ combinations.
As was noticed in Ref.\cite{Kulagin:1998wc},
the relative shadowing effect
for $F_3$ (or $q^-$) is enhanced by the factor of 2 of that of $F_2$ if
the real parts of effective scattering amplitudes vanish.
The presence of the real parts
leads to important interference effect thus reducing the shadowing effect \cite{KP04,KP14}.
}
We also remark that the result of averaging of the nucleon PDF
with nuclear spectral function
[the FMB effect, see the first two terms in the right side of Eq.(\ref{npdf})]
depends on the details of the $x$ dependence of considered PDF.
In particular, the combined FMB and OS correction at small $x$
is positive for the valence quark and negative for the sea quark and gluon PDFs.

In the intermediate region of $x\sim 0.1$
(which is usually referred to as \emph{antishadowing} region)
we observe interplay between different nuclear corrections. 
For the valence quark PDF ($C$-odd $q^-$ distribution) we find an enhancement
which is due to constructive interference in the multiple scattering effect between
the $C$-even and $C$-odd amplitudes in the $\delta q^\mathrm{coh}$ term.
The antishadowing effect on  the antiquark PDFs is significantly weaker
because of a partial cancellation between different contributions.
We also note that, unlike the $\bar u$ and $\bar d$,
the nuclear $\bar s$  in Fig.\ref{fig:npdfs} does not include  the MEC term. 

At large $x>0.2$ the nuclear corrections are dominated by the FMB and OS corrections.
For the valence quarks, those corrections form a pronounced `EMC-effect' shape at large $x$ \cite{KP04}.
However, as was mentioned above, the FMB correction crucially depends on the $x$ shape of the PDF
and for this reason is different for antiquark distribution.

\begin{figure}[htb]
\centering
\includegraphics[width=\textwidth]{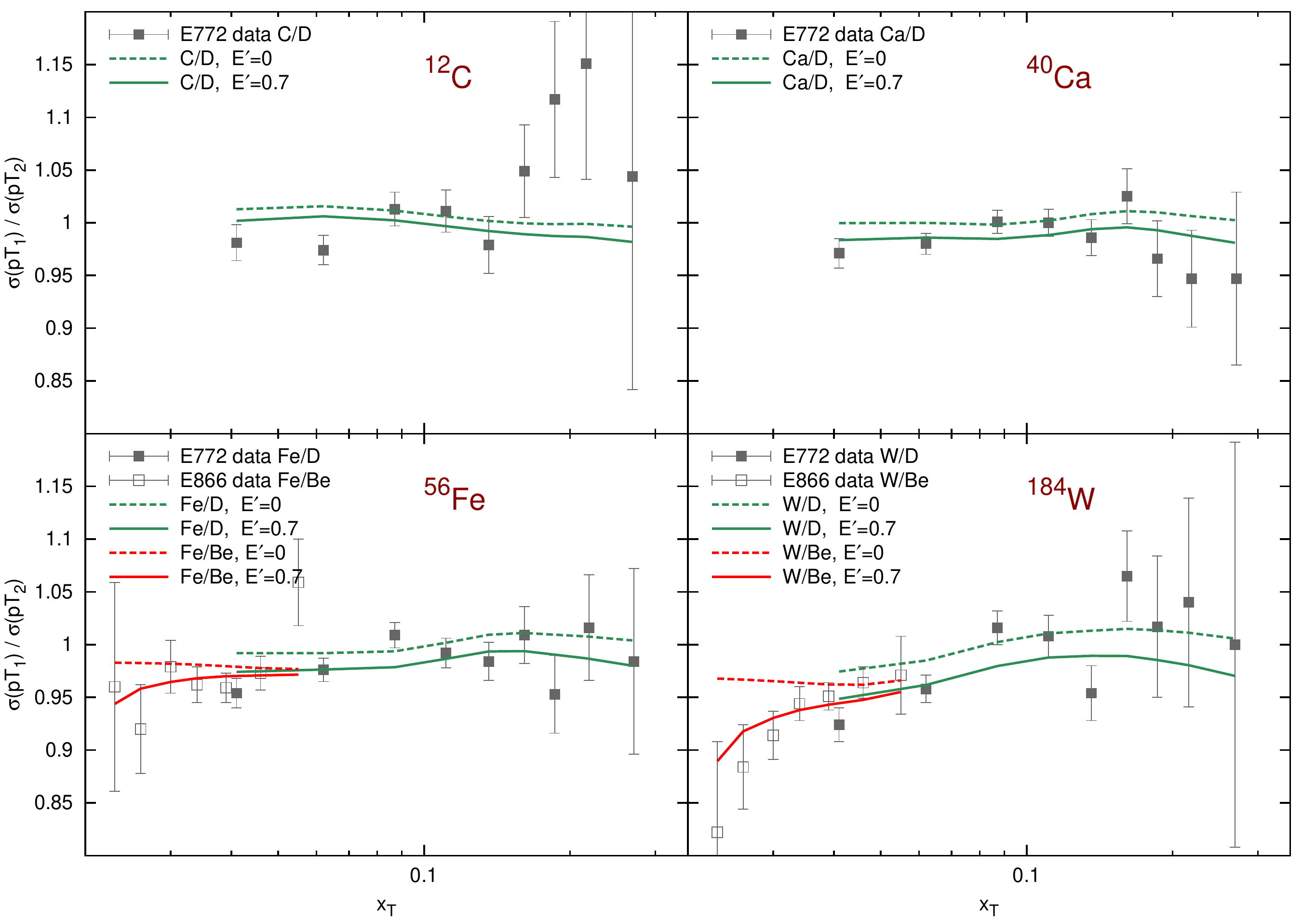}
\caption{%
(Color online)
Ratio of the DY reaction cross sections for different nuclei vs. $x_T$.
Data points are from the E772 experiment~\cite{E772} (solid boxes) and E866 experiment~\cite{E866} (open boxes).
The curves are the predictions of Ref.\cite{KP14} 
with (solid) and without (dashed) the projectile energy loss effect (see text for details).
\label{fig:dy}}
\end{figure}

Figure~\ref{fig:dy} displays the ratios of the Drell-Yan (DY) reaction cross sections
as a function of the target's Bjorken variable $x_T$, which were
calculated for the targets and the kinematics of E772~\cite{E772} and E866~\cite{E866} experiments.
The nuclear dependence of the DY process comes from two different sources: (i) the
modification of the nuclear target PDFs, and (ii) the initial state
interaction of the projectile particle (parton) within the nuclear environment of the target.

Note that for the E772 and E866 kinematics the DY nuclear ratios in the region
$x_T < 0.15$ are driven by the ratios of the antiquark distributions in the corresponding targets.
We recall that the lack of antishadowing in the DY data
at $x\sim 0.1$ was a long standing puzzle since the nuclear
binding should result in an
excess of nuclear mesons, which is expected to produce a marked enhancement in the nuclear
anti-quark distributions \cite{Bickerstaff:1985ax}.
As discussed above, our model predicts a significant cancellation of different nuclear effects
in the antiquark distribution in the region $x\sim 0.1$ in agreement with the nuclear DY data.
 
The projectile partons in the initial state can undergo multiple soft collisions
in nuclear medium and radiate gluons thus loosing the energy \cite{Bjorken:1982tu} before
annihilating with the (anti)quarks of the target and producing a dimuon pair.
In order to account for the quark energy loss effect,
let $E'$ be the parton energy loss in a nucleus per unit length.
If a projectile parton travels over the distance $L$ in the nuclear environment before annihilation,
then its
Bjorken variable $x_B$ changes by $E' L/E_p$, where $E_p$ is the energy of the projectile proton
(see, e.g., Ref.\cite{Garvey:2002sn}).
We estimate the average propagation length of the projectile partons in the uniform nuclear medium 
as $L=3R/4$, where $R$ is the nucleus radius.
Our analysis \cite{KP14} indicates that the E772 and E866 data favor the presence of
a moderate energy loss effect.
The solid curves in Fig.\ref{fig:dy} show our predictions with $E'=0.7$~GeV/fm.

The data from the E866 experiment is shifted towards lower values of target's $x_T$ and higher 
values of projectile's $x_B$ with respect to E772 data, as can be seen from Fig.\ref{fig:dy}.  
The kinematic coverage of E866 data is therefore focused on the region where both shadowing 
and energy loss effects become more prominent. The E866 data are consistent with the E772 data in the 
overlap region. Figure~\ref{fig:dy} shows that our predictions for the E866 and E772 kinematics are in 
good agreement data. We also note that the E866 and E772 experiments present somewhat different nuclear
ratios: $A/{}^2$H in the case of E772 and $A/^9$Be in the case of E866. For this reason the corresponding curves in
Fig.\ref{fig:dy} do not coincide in the E866--E772 overlap region.

\section{Summary}

In summary,
we presented the results of our studies~\cite{KP14} of nPDFs basing on an approach which was successfully applied to nuclear DIS in Refs.\cite{KP04,KP10}.
A number of mechanisms responsible for nuclear corrections on PDFs
in different regions of Bjorken $x$ was discussed.
We also discussed the application to the nuclear DY
process and found that our predictions on the ratio of the DY yields are in a good agreement
with the measurements by the E772 and E866 experiments.
We also studied the sensitivity of the nuclear DY reaction
to the energy loss effect of the projectile parton in nuclear environment,
the effect which is not included in phenomenological nPDF analyses.


The work of S.K. on the nuclear PDFs
was supported by the Russian Science Foundation grant No.~14-22-00161.
R.P. was supported by the grant DE-FG02-13ER41922 from the Department of Energy, USA.

\end{document}